\documentclass[journal=nalefd, manuscript=letter]{achemso}

\usepackage[version=3]{mhchem} 
\usepackage[T1]{fontenc}       
\usepackage{xcolor} 
\usepackage{amsthm,amsmath}
\usepackage{graphicx}

\author{Tengfei Liu}
\affiliation[Fuzhou University]
{Institute of Micro/Nano Devices and Solar Cells, School of Physics and Information Engineering, Fuzhou University, Fuzhou 350108, China}
\author{Xiyu Hong}
\affiliation[Tsinghua University]
{Department of Physics, State Key Laboratory of Low Dimensional Quantum Physics, Tsinghua University, Beijing 100084, China.}

\author{Zhe Li}
\affiliation[Chinese Academy of Sciences]
{Beijing National Research Center for Condensed Matter Physics, and Institute of Physics, Chinese Academy of Sciences, Beijing 100190, China}

\author{Shenzhong Chen}
\affiliation[CanSemi Technology Inc.]
{Technical department, CanSemi Technology Inc., Guangzhou 511363,China}

\author{Leyi Li}
\affiliation[Tsinghua University]
{Department of Physics, State Key Laboratory of Low Dimensional Quantum Physics, Tsinghua University, Beijing 100084, China.}

\author{Xin-Yi Tang}
\affiliation[Tsinghua University]
{Department of Physics, State Key Laboratory of Low Dimensional Quantum Physics, Tsinghua University, Beijing 100084, China.}

\author{Shuying Cheng}
\affiliation
{Institute of Micro/Nano Devices and Solar Cells, School of Physics and Information Engineering, Fuzhou University, Fuzhou 350108, China}%

\author{Yunfeng Lai}
\affiliation[Fuzhou University]
{Institute of Micro/Nano Devices and Solar Cells, School of Physics and Information Engineering, Fuzhou University, Fuzhou 350108, China}%

\author{Yonghai Chen}
\affiliation{Laboratory of Solid State Optoelectronics Information Technology, Institute of Semiconductors, Chinese Academy of Sciences, Beijing 100083, China}
\alsoaffiliation{College of Materials Science and Opto-Electronic Technology, University of Chinese Academy of Science, Beijing 101804, China.}

\author{Zhu Diao}
\affiliation[Maynooth University,]
{Department of Electronic Engineering and Maynooth International Engineering College,  Maynooth, Co. Kildare, Ireland.}

\author{Ke He}
\affiliation{Department of Physics, State Key Laboratory of Low Dimensional Quantum Physics, Tsinghua University, Beijing 100084, China.}

\author{Qikun Xue}
\affiliation[Tsinghua University]
{Department of Physics, State Key Laboratory of Low Dimensional Quantum Physics, Tsinghua University, Beijing 100084, China.}

\author{Jinling Yu}
\email{Jlyu@semi.ac.cn}
\affiliation[Fuzhou University]
{Institute of Micro/Nano Devices and Solar Cells, School of Physics and Information Engineering, Fuzhou University, Fuzhou 350108, China}

\title[]
  {Thickness-Induced Topological Phase Transition Investigated by Helicity Dependent Photocurrent in $\alpha$-Sn/CdTe(110)}

\abbreviations{TSs: topological surface states; SOC: spin-orbit coupling; TI: topological insulator; BL: bilayer; CPGE: circular photogalvanic effect; CPDE: circular photo drag effect; CPL: circularly polarized light}

\begin{document}







\newpage

\begin{abstract}

$\alpha$-Sn exhibits a rich topological phase diagram, yet experimental methods to tune and distinguish these phases remain limited. Here, we investigated the helicity-dependent photocurrent (HDPC) in $\alpha$-Sn films of varying thickness grown on CdTe(110) by molecular beam epitaxy. The HDPC of the 5 nm $\alpha$-Sn film shows an odd-function dependence on incident angle, whereas that of the 10 and 30 nm films exhibit an even-function dependence. Combined with high-resolution transmission electron microscopy (HR-TEM), point-group symmetry analysis, and first-principles calculations, it is revealed that a thickness-driven topological phase transition from a two dimensional (2D) to a three dimensional (3D) topological insulator occurs between 5 and 10 nm. These results demonstrate that HDPC serves as a sensitive diagnostic tool for topological phase transitions. The tunable electronic properties of $\alpha$-Sn(110) films enable thickness- and strain-mediated control of topological states, establishing a versatile platform for exploring emerging topological phenomena and developing spin-based devices.
\end{abstract}

\section{Introduction}
The $\alpha$-tin ($\alpha$-Sn) is a classic zero-gap semiconductor with a diamond cubic structure \cite{Hong2012,10.1063/5.0223869,HERMANOWICZ201676,Liu2022PairDW,MASSETTI2025102194,10.1063/5.0177343}. In the past decades, the $\alpha$-Sn has attracted renewed attention due to its rich topological phases.\cite{PhysRevB.98.115153,PhysRevLett.118.146402,10.1063/5.0084762,Zheng_2020, https://doi.org/10.1002/adfm.202008411,https://doi.org/10.1002/adma.202005909,PhysRevB.109.245135} Due to its pronounced sensitivity to strain, the epitaxial growth of $\alpha$-Sn on lattice-matched substrates has attracted significant research interest\cite{10.1126/science.aax3873,PhysRevMaterials.8.044202,PhysRevLett.113.256401,10.1007/s11467-020-0965-5,PhysRevB.92.081112}. Conventional substrates such as CdTe and InSb are widely used due to their close lattice matching with $\alpha$-Sn, enabling the deposition of high-quality thin films\cite{PhysRevLett.111.157205,POLACZYNSKI2024135,10.1063/5.0098585,Huang2017,10.1002/adfm.201802723}. Among these, epitaxial thick $\alpha$-Sn films (>10 nm) grown on CdTe (001) and CdTe (111) substrates have been extensively studied\cite{10.1116/6.0003564,PhysRevB.105.075109,10.1021/acsami.3c00323}. These orientations are favored for their relatively simple symmetries and well-established growth protocols, providing platforms to investigate strain-induced phase transitions.




However, research on $\alpha$-Sn films grown on unconventional crystal planes, such as (110), remains largely unexplored. The $\alpha$-Sn films grown on (110) plane exhibits distinct symmetry characteristics compared to the commonly studied (001) and (111) planes. In the diamond cubic structure of $\alpha$-Sn, the (001) plane is characterized by a square lattice with fourfold rotational symmetry\cite{PhysRevB.76.045302}, while the (111) plane exhibits hexagonal symmetry\cite{doi:10.1021/acs.jpclett.9b03538}. In contrast, the (110) plane is anisotropic, with a rectangular surface unit cell and reduced symmetry, which can lead to different strain configurations, electronic structures, and topological behaviors. The potential impact of these differences on the physical properties of $\alpha$-Sn thin films has not yet fully understood. Moreover, conventional characterization techniques predominantly rely on electrical or magnetic transport measurements to probe topological properties of materials. Such methods, however, suffer from an intrinsic ambiguity: they cannot definitively distinguish whether observed signals originate from bulk states or surface states. In contrast, helicity-dependent photocurrent (HDPC) measurements circumvent this limitation through symmetry constraints. By analyzing the incident-angle dependence of photocurrents, HDPC unambiguously discriminates between surface and bulk contributions, thereby serving as a powerful probe for topological phase transitions and interfacial spin-orbit coupling (SOC).\cite{10.1038/nnano.2011.214}


In this work, we report the successful epitaxial growth of high-quality $\alpha$-Sn thin films on a CdTe(110) substrate using molecular beam epitaxy (MBE) and utilize HDPC to investigate the thickness-driven topological phase transition of the $\alpha$-Sn films grown on CdTe(110) substrates. Pronounced differences in the incident angle dependence of HDPC are observed between the 5 nm and the thicker 10 nm and 30 nm (110)$\alpha$-Sn films, suggesting a topological phase transition as the thickness increases from 5 to 10 nm. Subsequently, the point symmetry of the $\alpha$-Sn/CdTe (110) films are analyzed, revealing distinct differences from (001) and (111) oriented $\alpha$-Sn films, attributed to their unique growth orientation. In addition to our experimental findings, by using first-principles calculations, we investigate the influence of the (110) crystal orientation on the electronic structure and topological properties of $\alpha$-Sn. The calculation results further confirm the occurrence of a thickness-driven topological phase transition from a two-dimensional (2D) to a three-dimensional (3D) topological insulator (TI) when the film thickness increases from 5 to 10 nm, which is distinct different from that observed in (001) and (111) $\alpha$-Sn films.


\section{Characterization of the $\alpha$-Sn/(110) CdTe thin films}
\begin{figure}
    \centering
    \includegraphics[width=0.8\textwidth]{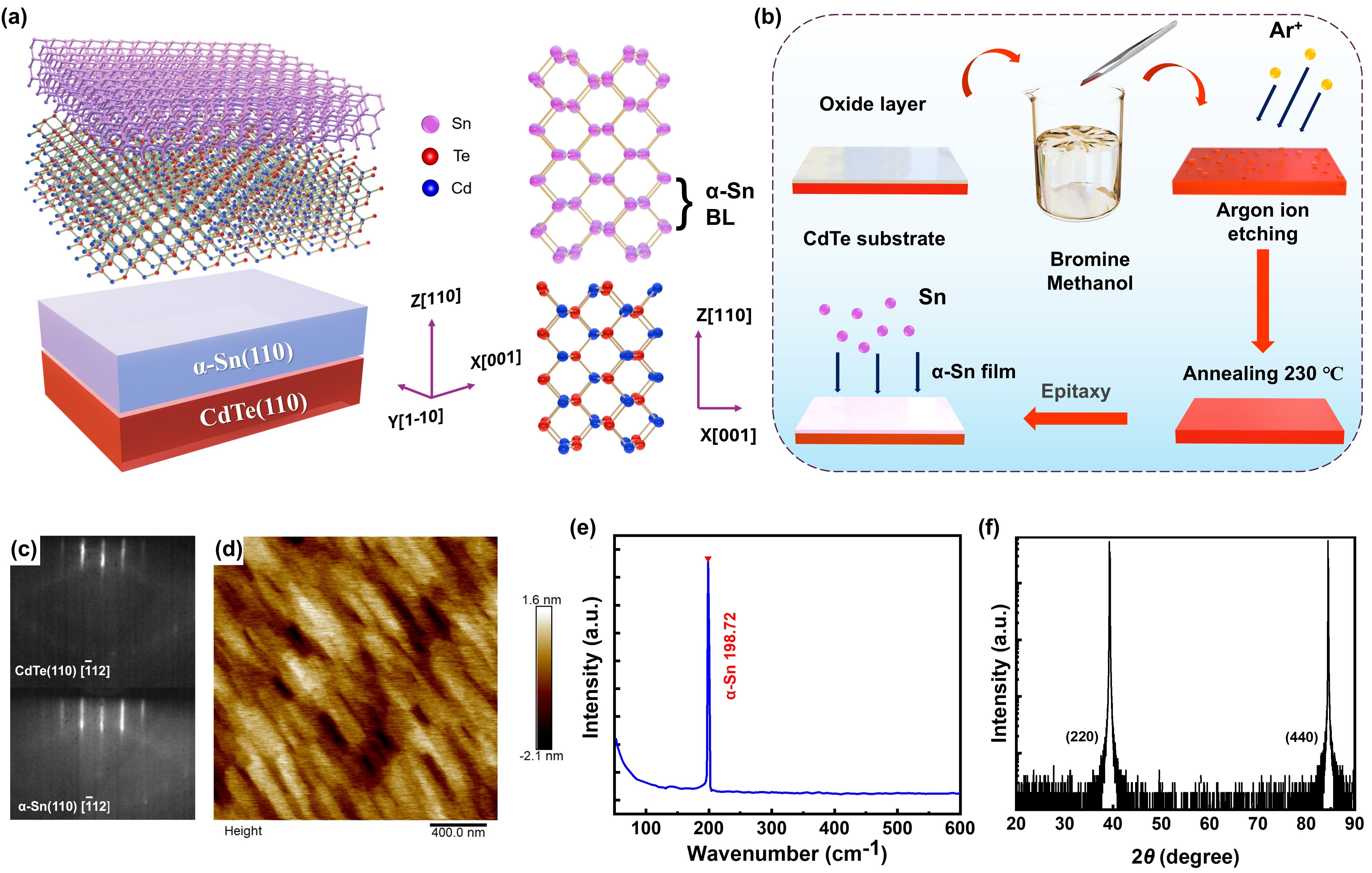}
    \caption{Growth and characterization of $\alpha$-Sn/CdTe(110) films. (a) Schematic drawing of the atomic structure and the cross-sectional views for the $\alpha$-Sn/CdTe(110) heterostructure. The pink, red, and blue spheres represent Sn, Te and Cd atoms, respectively. BL is bilayer of $\alpha$-Sn(110). (b) Fabrication flow diagram of the $\alpha$-Sn/CdTe(110) films. (c) RHEED patterns of the CdTe(110) substrate and the 5 nm $\alpha$-Sn(110) thin film along the [$\overline{1}$12] direction, indicating high quality of the epitaxial growth of $\alpha$-Sn(110) thin films. (d) Surface morphology of the 5 nm $\alpha$-Sn(110) film as observed by AFM. (e) Raman spectrum of the 5 nm $\alpha$-Sn(110) film with 532 nm excitation, confirming the absence of $\beta$-Sn signatures. (f) XRD pattern of a 5 nm $\alpha$-Sn/CdTe(110) thin film acquired at grazing incidence, demonstrating the crystallinity and phase purity of the heterostructure.
    }
    \label{figure1}
\end{figure}

The $\alpha$-Sn thin films with thicknesses of 5, 10, and 30 nm are grown on CdTe(110) substrates by MBE. Figure \ref{figure1}a-b show the schematic drawings of the three-dimensional and cross-sectional views of atomic structure for the $\alpha$-Sn films grown on CdTe(110) substrates, respectively. Here, bilayer (BL) serve as the thickness-calibration unit, where the lattice constant of 1 BL $\alpha$-Sn film is 4.63 \AA, representing a complete (110)-oriented unit cell. Figure \ref{figure1} c shows the reflection high-energy electron diffraction (RHEED) patterns of the CdTe(110) substrate and the $\alpha$-Sn(110)  thin film. It can be seen that the RHEED patterns display well-ordered stripes and Kikuchi lines following ion bombardment and annealing treatment of the CdTe substrate, confirming the presence of a high-quality substrate surface. Figure \ref{figure1} d shows the surface morphology of the $\alpha$-Sn(110)  film of 5 nm measured by atomic force microscopy (AFM), which demonstrates a smooth surface morphology, with a root mean square roughness of $R_q = 0.545$ nm. Furthermore, all samples, regardless of thickness, exhibit consistent surface roughness values below 1 nm, indicating comparable high quality to previously reported $\alpha$-Sn thin films grown on (111) and (001) planes. During the initial stages of Sn deposition, the RHEED patterns became diffuse. However, after 2 nm of Sn deposition, clear and sharp stripes re-emerged and remained consistent throughout the entire deposition process. 

Figure \ref{figure1}e shows the Raman spectrum of the 5 nm $\alpha$-Sn/CdTe(110) sample, which shows a prominent intensity peak at 198.72 cm$^{-1}$, confirming the absence of $\beta$-Sn in the sample and suggesting the phase purity. Figure \ref{figure1}f displays the X-ray diffraction (XRD) pattern of the 5 nm $\alpha$-Sn/CdTe(110) sample, which reveals a distinct $\alpha$-Sn peak coinciding with the substrate peak, due to the nearly identical lattice structures of the epitaxial thin film and the substrate. In addition, previous studies have shown that strain relaxation does not occur even in films several hundred nanometers thick, indicating that all our samples are subjected to the same strain.\cite{10.1116/6.0000756} Notably, no signals corresponding to the $\beta$-Sn phase are observed across the entire XRD spectrum, which is consistent with that observed in Raman spectrum, further confirming the phase purity.

\section{HDPC Photocurrent Mesurements} 
\begin{figure}
    \centering
    \includegraphics[width=0.8\linewidth]{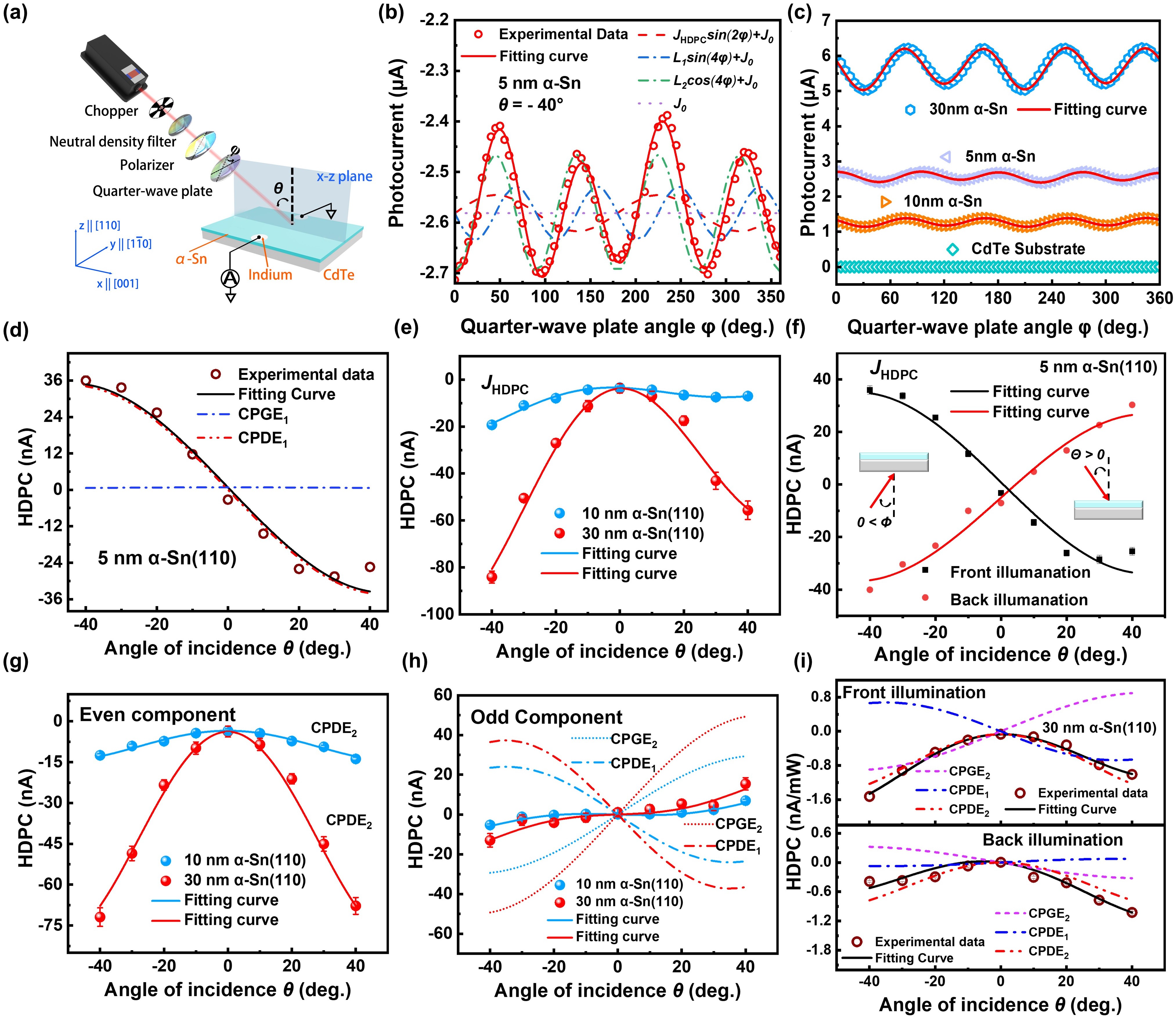}
    \caption{Experiment setup of HDPC and the incident-angle dependent HDPC of the $\alpha$-Sn/CdTe(110) films. (a) Schematic diagram for the HDPC measurement geometry. (b) Photocurrent as a function of the quarter wave plate angle $\varphi$ at an incident angle of -40$^\circ$ under an illumination of 1064 nm light. The red circles are the experimental data, and the red solid line is the fitting curve using eq \ref{eq:1}. The red dashed line indicates the photocurrent component $C\sin2\varphi + J_0$, and the blue and green dashed and dotted lines represent the photocurrent components of $L_1\sin4\varphi+ J_0$ and $L_2\cos4\varphi+ J_0$, respectively. The blue dashed line denotes the polarization independent photocurrent $J_0$. (c) Dependence of the photocurrent on the quarter-wave plate angle $\varphi$ measured at an incident angle $\theta$ of -40$^\circ$ for the 5, 10, 30 nm $\alpha$-Sn/CdTe films and CdTe(110) substrate. (d) Dependence of the HDPC on the incident angle in the 5 nm $\alpha$-Sn(110)  film. (e) Dependence of the HDPC on the incident angle in the 10 and 30 nm $\alpha$-Sn(110) films. (f) Dependence of the HDPC for the 5 nm $\alpha$-Sn(110) film under the front and back illuminations of 1064 nm light. The solid lines are the fitting curve by using eq 2. (g-h) Even- and odd-function components of the incident angle extracted from the total HDPC current of the 10 and 30 nm $\alpha$-Sn(110)  films, respectively. The solid lines are the fitting curves by using CPDE$_2$ and CPGE$_2$+CPDE$_1$, respectively. (i) Dependence of the HDPC on the incident angle under the front and back illuminations for the 30 nm $\alpha$-Sn(110) film. The solid lines represent the fitting curves by using CPDE$_2$+CPDE$_1$+CPGE$_2$.}
    \label{fig:2}
\end{figure}

The HDPC of the $\alpha$-Sn/CdTe(110) films with different thicknesses was measured, and the experiment setup is shown in Figure \ref{fig:2}a (See Figure S1 for more detailed information). The laser incident plane lies in the $x$-$z$ plane, and the photocurrent following along the $y$ direction is collected. Figure~\ref{fig:2}b shows the typical photocurrent as a function of the quarter-wave plate angle $\varphi$ of the 5 nm $\alpha$-Sn/CdTe (110) film  under illumination of 1064 nm light with a power of 55 mW. As we can see, the photocurrent $J$ varies with the polarization state of the light, which can be generally expressed by eq \ref{eq:1}\cite{PhysRevB.93.125434}:
\begin{equation}\label{eq:1}
J=J_{\rm{HDPC}}\sin2\varphi+L_\textit{1}\sin4\varphi+L_\textit{2}\cos4\varphi+J_\textit{0},
\end{equation}
where $\varphi$ is the angle between the polarization direction of the incident light and the optical axis of the quarter-wave plate, and $J_{\rm{HDPC}}$ represents the HDPC current. $L_1$ and $L_2$ are the photocurrent induced by linearly polarized light, and $J_0$ is the polarization-independent photocurrent, which is mainly generated by the thermoelectric effect, photovoltaic effect or Dember effect\cite{PhysRevB.94.161405,PhysRevB.100.235108}.

Figure \ref{fig:2}c shows dependence of the photocurrent on the quarter-wave plate angle $\varphi$ for the $\alpha$-Sn/CdTe(110) films with different thicknesses and the CdTe(110) substrate. It can be seen that the photocurrent of the CdTe(110) substrate is negligibly small compared with that of the $\alpha$-Sn films, confirming that the HDPC of the $\alpha$-Sn/(110) CdTe samples arises almost entirely from the $\alpha$-Sn films. Figure \ref{fig:2}d and g show the dependence of the HDPC current on the incident angle of light for the $\alpha$-Sn/CdTe(110)  films with 5, 10 and 30 nm. Surprisingly, the HDPC of the 5 nm sample shows an odd function on the incident angles, while that of the 10 and 30 nm samples almost display an even function on the incident angles. Generally, thin films epitaxially grown on identical substrates exhibit the same point-group symmetry, so their HDPC behavior would be expected to exhibit similar characteristics, with differences primarily in magnitude. Therefore, the different incident angle dependence of HDPC for the (110)$\alpha$-Sn films with different thicknesses may indicate that the point-group symmetry has changed, which may be induced by the topological phase transition as the increase of the film thickness. 

To determine the allowed forms of the photocurrent, we analyze the point-group symmetry of the samples. The $\alpha$-Sn crystallizes in the diamond structure with $Fd$$\overline{3}$$m$ symmetry (space group No. 227)\cite{PhysRevMaterials.6.113601}. We performed the cross sectional high-resolution transmission electron microscopy (HR-TEM) measurements to determine the strain of $\alpha$-Sn(110)  films, and thus to obtain their point group symmetry. 

\begin{figure}
    \centering
    \includegraphics[width=0.8\linewidth]{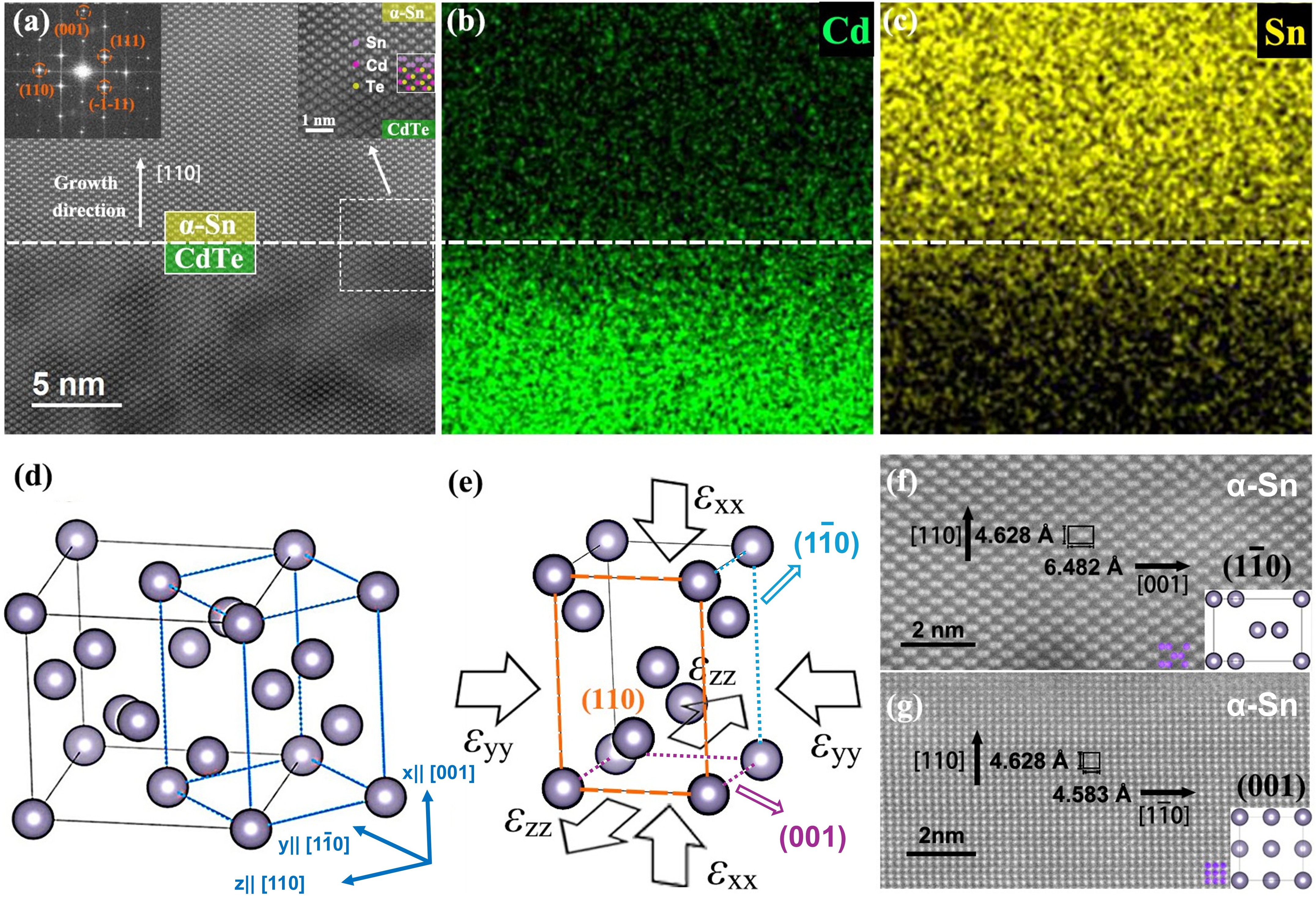}
\caption{
TEM image and strain analysis. (a) HR-TEM image of the interface between the $\alpha$-Sn thin film and the CdTe(110) substrate, accompanied by fast Fourier transform (FFT) patterns that confirm the epitaxial growth orientation. The inset highlights the atomic alignment along the [110] growth direction. (b, c) Elemental mapping of Cd and Sn obtained via energy-dispersive X-ray spectroscopy (EDS), showing a sharp interface between the $\alpha$-Sn thin film and the CdTe(110) substrate. Green and yellow dots represent the Cd and Sn atomic signals, respectively. (d) Crystal structure of $\alpha$-Sn, with the (110) plane depicted as a blue rectangular unit cell, while Sn atoms are represented by purple spheres across all panels. (e) Schematic diagram illustrating the strain components $\varepsilon_{xx}$, $\varepsilon_{yy}$, and $\varepsilon_{zz}$ applied to the $\alpha$-Sn(110), with arrows indicating the possible strain directions. The orange dashed square highlights the (110) plane. (f) Atomic-resolution HR-TEM image of $\alpha$-Sn(110), displaying lattice spacings of 4.628 Å and 6.482 Å corresponding to the [001] and [1$\bar{1}$0] directions, respectively. (g) A similar HR-TEM image to (f), but showing a different side view, with lattice spacings of 4.628 Å and 4.583 Å corresponding to the [110] and [0$\bar{1}$0] directions, respectively.
}
\label{fig:3}
\end{figure}

Figure \ref{fig:3}a shows the HR-TEM image of the interface between the 5 nm $\alpha$-Sn thin film and the CdTe(110) substrate, which reveals a continuous Sn-CdTe lattice with a straight, dislocation-free interface, demonstrating the excellent quality of the film. Figure \ref{fig:3}b and c show the elemental mapping of Cd and Sn obtained by energy-dispersive X-ray spectroscopy (EDS). It can be seen that the Cd atoms (green) are well confined to the CdTe substrate, exhibiting a well-defined and atomically sharp interface with adjacent Sn (yellow). The schematic of the $\alpha$-Sn unit cell and substrate-induced stress on $\alpha$-Sn(110) is schematically illustrated in Figure~\ref{fig:3}f and e. Since the unit cell of the CdTe(110) plane exhibits a rectangular shape with the lattice constant smaller than that of the $\alpha$-Sn, the [001] ($x$) and [1$\bar{1}$0] ($y$) directions of the $\alpha$-Sn(110)  film will experience compressive stress, and the [110] ($z$) direction will be under tensile stress, as shown in Figure~\ref{fig:3}e. Figure \ref{fig:3}f and g show the HR-TEM images of (1$\bar{1}$0) and the (001) crystal planes, respectively, for the 5 nm $\alpha$-Sn thin film grown on CdTe(110) substrate, which clearly demonstrate the presence of the substrate-induced strain on the $\alpha$-Sn thin film. Specifically, the lattice constants along the [001] and [1$\bar{1}$0] axis contract from 6.489  to 6.482 \AA\  and from 9.177  to 9.166 \AA, respectively,  reflecting compressive strain. The lattice constant along [110] axis expands from 9.177 to 9.255 \AA, corresponding to a tensile strain of approximately 0.86\%. This anisotropic strain distribution results in a deformation of the cubic crystal system of $\alpha$-Sn, transforming it into a monoclinic crystal system. 


Due to the substrate-induced strain, the point-group symmetry of the bulk states is reduced from $O_h$ to $C_{2v}$. For the interface or surface states, the point-group symmetry is reduced from  $C_{2v}$ to $C_s$, due to lacking of the two-fold axis and the in-plane mirror. The related tensor of crystal point group was analyzed and deduced, please refer to notes S2 and S3 of the supplementary material for detailed information.  

There are two origins for the HDPC, i.e., CPGE and CPDE. Therefore, the HDPC of $C_{2v}$ and $C_{s}$ point-group symmetries can be expressed by the following equations under the geometry shown in Figure \ref{fig:2}a:

\begin{equation}\label{eq:9}
    J_{HDPC}^{C_{2v}}=\gamma_{yz}P_{circ}|E_0|^2t_pt_s\cos{\theta}+{\widetilde{T}_{yxz}{{q_zP}_{circ}\left|E_0\right|}^2t_pt_s \sin\theta \cos\theta},
\end{equation}
and 
\begin{equation}\label{eq:10}
\begin{gathered}
    J_{HDPC}^{C_{s}}=\gamma_{yx}P_{circ}|E_0|^2t_pt_s\sin{\theta}+{\widetilde{T}_{yxz}{{q_zP}_{circ}\left|E_0\right|}^2t_pt_s\sin\theta \cos\theta}+{\widetilde{T}_{yxx}{{q_xP}_{circ}\left|E_0\right|}^2t_pt_s\sin^2\theta}.
\end{gathered}
\end{equation}
Here, $P_{\rm{circ}}$ is the light helicity, and $t_p$ and $t_s$ are the transmission coefficients for linear p- and s-polarized light, respectively. $\theta$ is the angle of refraction, defined by sin$\theta$=sin$\theta_0$/$\sqrt{\varepsilon}$, with $\varepsilon$ and $\theta_0$ being the dielectric constant of $\alpha$-Sn and the incident angle of the light, respectively. $q$ is the photon wave vector. The first and second terms of eq~\ref{eq:9} is denoted as $\rm{CPGE_1}$ and $\rm{CPDE_1}$, respectively, and the first and last terms of eq~\ref{eq:10} is denoted as $\rm{CPGE_2}$ and $\rm{CPDE_2}$, respectively. 


The HDPC of the 5 nm sample shows an odd function of the incident angles, which may be contributed by $\rm{CPDE_1}$ or by the $\rm{CPGE_2}$ of the surface states or interface states with $C_{s}$ symmetry. In contrast, the 10 and 30 nm samples show a dominant even component with a minimum at zero incidence angle consistent only with the $\sin^2\theta$ term permitted by $C_{s}$ symmetry, thereby confirming that the HDPC is dominated by the surface states or the interface states in the 10 and 30 nm $\alpha$-Sn (110) films. $\rm{CPDE_1}$ shows a $\sin2\theta$ dependence while $\rm{CPGE_2}$ shows a $\sin\theta$ dependence, which provides us a method to distinguish which current plays the dominant role in the HDPC of the 5 nm sample. Since the HDPC of the 5 nm sample shows a better linear dependence on $\sin2\theta$, as shown in Figure S2 in the supplementary material, the HDPC of the 5 nm sample can be mainly attributed to the $\rm{CPDE_1}$. To further confirm our inference, we perform the HDPC measurements for the 5 nm sample under the front and back illuminations, and the result is shown in Figure \ref{fig:2}f. It can be seen that, the sign of the HDPC reverses when the incident light changes from the front illumination to the back illumination, which is due to the reversal of the direction of the light propagation $q_z$ for the back illumination. The HDPC of the 5 nm sample under the front and back illuminations can be well explained by $\rm{CPDE_1}$, thus confirming our inference. Then, we fit the incident angle dependent HDPC current of the 5 nm $\alpha$-Sn with eq~2, and the fitting result is shown by the solid line in Figure~\ref{fig:2}d, which shows a very good agreement with the experimental data. The dotted and dashed-dotted lines indicate the $\rm{CPDE_1}$ and $\rm{CPGE_1}$ current obtained by fitting, suggesting a dominant contribution of $\rm{CPDE_1}$.




It can be seen from Figure~\ref{fig:2}e that when the thickness of the $\alpha$-Sn (110) film is larger than 10 nm, the surface states or the interface states play the dominant role in the HDPC, which may indicate the presence of the surface or interface states with strong SOC when the thickness is thicker than 10 nm. It is obvious that the HDPC in the 10 nm and 30 nm samples contain both even- and odd-function components with respect to the incident angles. By using $[J_{\rm{HDPC}}(+\theta)+J_{\rm{HDPC}}(-\theta)]/2$ and $[J_{\rm{HDPC}}(+\theta)-J_{\rm{HDPC}}(-\theta)]/2$, we can extract the even- and odd-function components of the incident angle from the total HDPC current, as shown in Figure~\ref{fig:2}g and h, respectively. The solid lines in Figure~\ref{fig:2}g and h are the fitting curves by using $\rm{CPDE_2}$ and $\rm{CPGE_2}+\rm{CPDE_1}$, respectively, which show very good agreements with the experimental data. The dotted and dashed-dotted lines in Figure~\ref{fig:2}h represent the $\rm{CPGE_2}$ and $\rm{CPDE_1}$ obtained by fitting, respectively. 

To investigate the origin of the surface or interface states, we have performed the HDPC measurements for the 30 nm sample under the front and back illuminations, and the results are shown in Figure \ref{fig:2}i. It can be seen that, as the light incidence changes from front illumination to the back illumination, the sign of the $\rm{CPGE_2}$ and $\rm{CPDE_1}$ reverses, while that of the $\rm{CPDE_2}$ remains unchanged. The sign reversal of $\rm{CPGE_2}$ may indicate the presence of the bottom topological surface states (TSS) of 3D TI, implying that the material undergoes a topological phase transition into a 3D TI phase. Thus, when the light incident changes from the front illumination to the back illumination, the dominant contribution of the $\rm{CPGE_2}$ is changed from top TSS to the bottom TSS. Given that the tensor  $\widetilde{T}_{yxx}$ exhibits opposite signs for the top and bottom TSS of a 3D TI, the unchanged sign of $\rm{CPDE_2}$ under both the front and back illuminations indicates that the $\rm{CPDE_2}$ in the 30 nm sample is dominated exclusively by a certain surface state--either the top or bottom surface state of the 3D TI. The larger $\rm{CPDE_2}$ measured under the front illumination than that under the back illumination may indicate that the $\rm{CPDE_2}$ is dominated by the top TSS of the 3D TI. The $\rm{CPDE_1}$ arises from two possible origins: the bulk states of the $C_{2v}$ symmetry or the TSS of the 3D TI $\alpha$-Sn(110). If the $\rm{CPDE_1}$ signal is dominated by bulk states, its sign reversal can be attributed to the variation in the wavevector  $q_z$. On the other hand, if TSS are the primary contributors, the sign reversal of $\rm{CPDE_1}$ indicates that the current is derived from a specific TSS—either the top or bottom TSS—under both front and back illumination conditions. This conclusion stems from the following two reasons: first, the topological tensor $\widetilde{T}_{yxz}$ exhibits opposite signs for the top and bottom TSS; second, the sign of $q_z$ reverses when the light incidence switches from front to back illumination. Therefore, the HDPC measured under the front and back illuminations demonstrates that when the film thickness reaches or exceeds 10 nm, the $\alpha$-Sn thin film epitaxially grown on CdTe(110) substrate undergoes a topological phase transition into a 3D TI phase, exhibiting both top and bottom surface states.





It can be seen from Figure~\ref{fig:2}g and h that as the film thickness increases from 10 to 30 nm, the intensity of $\rm{CPGE_2}$, $\rm{CPDE_1}$ and $\rm{CPDE_2}$ all increases. This phenomenon can be attributed to the weaker hybridization effect between the top and bottom surface states with the increase of the film thickness of 3D TI $\alpha$-Sn.\cite{d5xc-bvmx,WOS:000280559300013} Besides, if the $\rm{CPGE_2}$, $\rm{CPDE_1}$ and $\rm{CPDE_2}$ are dominated by the top TSS of the 3D TI  $\alpha$-Sn(110) under the front illumination, the light reaches at the bottom TSS will decrease as the increase of the film thickness, which will reduce the photocurrent of the bottom TSS. Considering that the sign of the HDPC in the bottom TSS is opposite to that of the top TSS, the reduction of the contribution of the bottom TSS will enhance the net HDPC current.

\begin{figure}[H]
    \centering
    \includegraphics[width=0.8\linewidth]{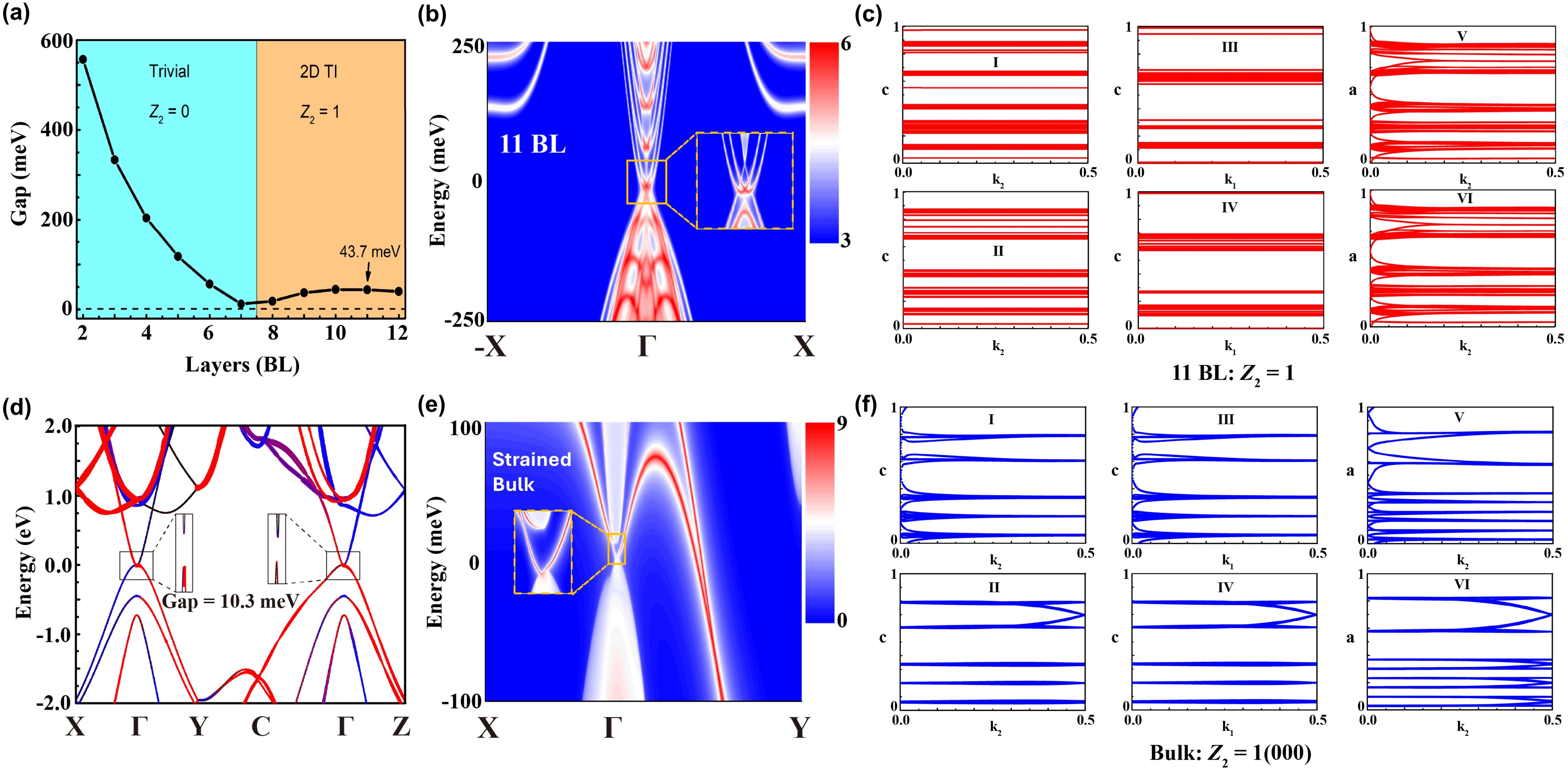}
    \caption{Theoretical calculation results based on first-principles calculations for the thin-film and strained bulk phase of $\alpha$-Sn(110)  under compressive strain. (a) The evolutions of gaps versus (vs.) the values of BL. Black and red lines stand for the $\Gamma$-gap and full gap values respectively, meanwhile light-blue and orange zones are corresponding to topological trivial (2D, $Z_2$ = 0) and nontrivial (2D, $Z_2$ = 1) phases. The horizontal dashed line denotes the position of zero gap. (b) LDOS pattern of a 11-BL $\alpha$-Sn film projected onto the Y edge. The zoomed-in pattern around the Dirac cone is depicted within the yellow frame. (c) Wannier Charge Center evolutions of 11-BL $\alpha$-Sn film along the six reciprocal planes involving $k_1$ = 0 (denoted as I), $k_1$ = 0.5 (denoted as II), $k_2$ = 0 (denoted as III), $k_2$ = 0.5 (denoted as IV), $k_3$ = 0 (denoted as V) and $k_3$ = 0.5 (denoted as VI), which provides 2D-$Z_2$ index as 1. (d) Orbital-projected band structure of strained bulk $\alpha$-Sn along the X (0.5, 0.0, 0.0) - $\Gamma$ (0.0, 0.0, 0.0) - Y (0.0, 0.5, 0.0) - C (0.5, 0.5, 0.0) - $\Gamma$ (0.0, 0.0, 0.0) - Z (0.0, 0.0, 0.5) path. Two zoom-in band structures near the $\Gamma$ point are displayed along X-$\Gamma$-Y and C-$\Gamma$-Z paths. (e) LDOS pattern of strained bulk $\alpha$-Sn cut along [110] semi-infinite surface. (f) Wannier Charge Center evolutions of strained bulk $\alpha$-Sn depicted similar with subfigure (c), which provides a 3D-$Z_2$ index as 1(000).}
    \label{fig:4}
\end{figure}

To further confirm our inference, we have performed the first-principles calculations for $\alpha$-Sn thin films grown on CdTe(110) substrates. Figure~\ref{fig:4}a summarizes the evolution of the topological phases and $\Gamma$-gaps of $\alpha$-Sn(110) with even BL thickness by utilizing  Heyd-Scuseria-Ernzerhof (HSE06) functional. Owing to the inversion-symmetry conservations, we can adopt one Z$_2$ index to describe and classify their 2D topological characters. In Figure~\ref{fig:4}a, the cyan and orange zones stand for Z$_2$ = 0 and 1, corresponding to topological trivial and nontrivial phase, respectively. Figure \ref{fig:4}b shows the Local-density-of-state (LDOS) pattern of 11-BL $\alpha$-Sn film projected onto the Y edge. The insert image shows Dirac cone, which characterizes the one-dimensional edge states of a 2D TI. Figure~\ref{fig:4}c shows the Wannier Charge Center (WCC) evolutions of 11-BL $\alpha$-Sn film along the six reciprocal planes, which gives $Z_2$=1 for the 2D system. It can be seen that, a topological phase transition from a trivial semiconductor (or semimetal) to 2D TI appears between 7 and 8 BL, and the $\alpha$-Sn(110) thin film with 11 BL, corresponding to 5 nm, is a 2D TI.  
Figure \ref{fig:4}d shows the orbital-projected band structure of strained bulk $\alpha$-Sn under compressive strain observed in our experiments, i.e., the [001] and [1$\bar{1}$0] directions are under compressive strains of 0.11$\%$ and 0.12$\%$, respectively, and the [110] direction is under a tensile strain of 0.86$\%$. Here, the red, blue and purple line are related to Sn-5$p_z$, Sn-5$p_x$ and Sn-5$s$ orbital contributions, respectively. The inset figure indicates an opened band gap of about 10.3 meV. Figure \ref{fig:4}e shows the LDOS pattern cut along (110) semi-infinite surface of strained bulk $\alpha$-Sn under the same compressive strain as that in Figure \ref{fig:4}d. The LDOS increases across the color gradient from blue to white to red, with the red curves specifically corresponding to the TSS. The whole pattern is depicted along X (0.5, 0.0) - $\Gamma$ (0.0, 0.0) - Y (0.0, 0.5) path. The zoomed-in LDOS pattern reveals a detailed distribution near the TSS. The WCC evolutions along the six reciprocal planes shown in Figure \ref{fig:4}f provides $Z_2$ as 1(000). Therefore, the thicker $\alpha$-Sn (110) film under in-plane compressive strain can be identified as a 3D TI. This is similar to the previously reported theoretical result,\cite{Huang2017} where bulk $\alpha$-Sn undergoes a transition into a 3D TI under a compressive strain along the [001] direction and in-plane (001) tensile strain. The theoretical calculations based on first-principles indicate that the 5 nm $\alpha$-Sn/CdTe(110) is a 2D TI, and thicker film (with the thickness at least larger than 12 BL) can be regarded as a 3D TI. The theoretical calculations are consistent with our experimental observation that a topological phase transition to a 3D TI appears between 5 and 10 nm, thus confirming our inference.

In summary, thickness-induced topological phase transition of in-plane strained $\alpha$-Sn thin films grown on CdTe(110) substrates have been systematically investigated by HDPC. The high quality of the $\alpha$-Sn thin films are characterized by RHEED, Raman, XRD and HR-TEM. Combining with the theoretical calculations based on first-principles and the HDPC measurement under the front and back illuminations, it is revealed that the 5 nm $\alpha$-Sn/CdTe(110)  film is a 2D TI, with the HDPC dominated by $\rm{CPDE_1}$.  When the film thickness is increased to 10 nm, the $\alpha$-Sn/CdTe (110) film undergoes a phase transition to a 3D TI. This triggers the emergence of topological surface states at the top and bottom (or interface) regions, causing the HDPC to shift from being dominated by an odd-function angular dependence to an even-function dependence. This study represents a significant step forward in understanding the role of crystal orientation and film thickness in shaping the physical properties of $\alpha$-Sn thin films. The distinct properties of (110)-oriented $\alpha$-Sn films open up new possibilities for engineering strain and electronic structures in $\alpha$-Sn, offering a versatile platform for studying novel topological phenomena and developing advanced topological functional devices.



\begin{acknowledgement}

The work was supported by the National Natural Science Foundation of China (No.~62074036, No.~61674038, No.~11574302), Foreign Cooperation Project of Fujian Province (2023I0005), Open Research Fund Program of the State Key Laboratory of Low-Dimensional Quantum Physics (No. KF202108), National Key Research and Development Program (2016YFB-0402303), and the Foundation of Fujian Provincial Department of Industry and Information Technology of China (Grant Nos. 82318075).

\end{acknowledgement}

\begin{suppinfo}
Experimental setup of the HDPC measurement, symmetry analysis for strained bulk state of $\alpha$-Sn (110) films,  symmetry analysis for the interface or surface state of $\alpha$-Sn (110) films, and distinguishment between CPGE$_2$ and CPDE$_1$.

\end{suppinfo}


\bibliography{achemso-demo}

\providecommand{\latin}[1]{#1}
\makeatletter
\providecommand{\doi}
  {\begingroup\let\do\@makeother\dospecials
  \catcode`\{=1 \catcode`\}=2 \doi@aux}
\providecommand{\doi@aux}[1]{\endgroup\texttt{#1}}
\makeatother
\providecommand*\mcitethebibliography{\thebibliography}
\csname @ifundefined\endcsname{endmcitethebibliography}
  {\let\endmcitethebibliography\endthebibliography}{}
\begin{mcitethebibliography}{37}
\providecommand*\natexlab[1]{#1}
\providecommand*\mciteSetBstSublistMode[1]{}
\providecommand*\mciteSetBstMaxWidthForm[2]{}
\providecommand*\mciteBstWouldAddEndPuncttrue
  {\def\EndOfBibitem{\unskip.}}
\providecommand*\mciteBstWouldAddEndPunctfalse
  {\let\EndOfBibitem\relax}
\providecommand*\mciteSetBstMidEndSepPunct[3]{}
\providecommand*\mciteSetBstSublistLabelBeginEnd[3]{}
\providecommand*\EndOfBibitem{}
\mciteSetBstSublistMode{f}
\mciteSetBstMaxWidthForm{subitem}{(\alph{mcitesubitemcount})}
\mciteSetBstSublistLabelBeginEnd
  {\mcitemaxwidthsubitemform\space}
  {\relax}
  {\relax}

\bibitem[Hong \latin{et~al.}(2012)Hong, Cha, Kong, and Cui]{Hong2012}
Hong,~S.~S.; Cha,~J.~J.; Kong,~D.; Cui,~Y. Ultra-low carrier concentration and
  surface-dominant transport in antimony-doped {B}i$_2${S}e$_3$ topological
  insulator nanoribbons. \emph{Nat. Commun.} \textbf{2012}, \emph{3}, 757\relax
\mciteBstWouldAddEndPuncttrue
\mciteSetBstMidEndSepPunct{\mcitedefaultmidpunct}
{\mcitedefaultendpunct}{\mcitedefaultseppunct}\relax
\EndOfBibitem
\bibitem[Li \latin{et~al.}(2025)Li, Ding, Yao, Fan, Liu, Chen, Zhou, Lu, and
  Chen]{10.1063/5.0223869}
Li,~B.; Ding,~Y.; Yao,~J.; Fan,~X.; Liu,~G.; Chen,~Y.-B.; Zhou,~J.; Lu,~H.;
  Chen,~Y.-F. Chiral anomaly in doped $\ensuremath{\alpha}$-Sn films by Fermi
  level tuning. \emph{Appl. Phys. Lett.} \textbf{2025}, \emph{126},
  093101\relax
\mciteBstWouldAddEndPuncttrue
\mciteSetBstMidEndSepPunct{\mcitedefaultmidpunct}
{\mcitedefaultendpunct}{\mcitedefaultseppunct}\relax
\EndOfBibitem
\bibitem[Hermanowicz and Radny(2016)Hermanowicz, and Radny]{HERMANOWICZ201676}
Hermanowicz,~M.; Radny,~M. Topological electronic states of bismuth selenide
  thin films upon structural surface defects. \emph{Computational Materials
  Science} \textbf{2016}, \emph{117}, 76--82\relax
\mciteBstWouldAddEndPuncttrue
\mciteSetBstMidEndSepPunct{\mcitedefaultmidpunct}
{\mcitedefaultendpunct}{\mcitedefaultseppunct}\relax
\EndOfBibitem
\bibitem[Liu \latin{et~al.}(2022)Liu, Wei, He, Zhang, Wang, and
  Wang]{Liu2022PairDW}
Liu,~Y.; Wei,~T.; He,~G.; Zhang,~Y.; Wang,~Z.; Wang,~J. Pair density wave state
  in a monolayer high-T$_c$ iron-based superconductor. \emph{Nature}
  \textbf{2022}, \emph{618}, 934--939\relax
\mciteBstWouldAddEndPuncttrue
\mciteSetBstMidEndSepPunct{\mcitedefaultmidpunct}
{\mcitedefaultendpunct}{\mcitedefaultseppunct}\relax
\EndOfBibitem
\bibitem[Massetti \latin{et~al.}(2025)Massetti, Crosta, {Le Mardelé},
  Mohelský, Martella, Molle, Orlita, Grazianetti, and
  Pezzoli]{MASSETTI2025102194}
Massetti,~C.; Crosta,~C.; {Le Mardelé},~F.; Mohelský,~I.; Martella,~C.;
  Molle,~A.; Orlita,~M.; Grazianetti,~C.; Pezzoli,~F. Quantum confinement
  effects in the topological Dirac semimetal $\ensuremath{\alpha}$-Sn on
  InSb(111). \emph{Matter} \textbf{2025}, 102194\relax
\mciteBstWouldAddEndPuncttrue
\mciteSetBstMidEndSepPunct{\mcitedefaultmidpunct}
{\mcitedefaultendpunct}{\mcitedefaultseppunct}\relax
\EndOfBibitem
\bibitem[Inagaki \latin{et~al.}(2024)Inagaki, Ishihara, Hotta, Seki, Takeda,
  Ishida, Ootsuki, Kawasaki, Fujimori, Tanaka, Anh, and
  Kobayashi]{10.1063/5.0177343}
Inagaki,~K.; Ishihara,~K.; Hotta,~T.; Seki,~Y.; Takeda,~T.; Ishida,~T.;
  Ootsuki,~D.; Kawasaki,~I.; Fujimori,~S.-I.; Tanaka,~M.; Anh,~L.~D.;
  Kobayashi,~M. Allotropic transition of Dirac semimetal $\alpha$-Sn to
  superconductor $\beta$-Sn induced by focused-ion-beam irradiation.
  \emph{Appl. Phys. Lett.} \textbf{2024}, \emph{124}, 021602\relax
\mciteBstWouldAddEndPuncttrue
\mciteSetBstMidEndSepPunct{\mcitedefaultmidpunct}
{\mcitedefaultendpunct}{\mcitedefaultseppunct}\relax
\EndOfBibitem
\bibitem[de~Coster \latin{et~al.}(2018)de~Coster, Folkes, Taylor, and
  Vail]{PhysRevB.98.115153}
de~Coster,~G.~J.; Folkes,~P.~A.; Taylor,~P.~J.; Vail,~O.~A. Effects of
  orientation and strain on the topological characteristics of
  CdTe/$\ensuremath{\alpha}$-Sn quantum wells. \emph{Phys. Rev. B}
  \textbf{2018}, \emph{98}, 115153\relax
\mciteBstWouldAddEndPuncttrue
\mciteSetBstMidEndSepPunct{\mcitedefaultmidpunct}
{\mcitedefaultendpunct}{\mcitedefaultseppunct}\relax
\EndOfBibitem
\bibitem[Xu \latin{et~al.}(2017)Xu, Chan, Chen, Chen, Wang, Dejoie, Wong,
  Hlevyack, Ryu, Kee, Tamura, Chou, Hussain, Mo, and
  Chiang]{PhysRevLett.118.146402}
Xu,~C.-Z.; Chan,~Y.-H.; Chen,~Y.; Chen,~P.; Wang,~X.; Dejoie,~C.; Wong,~M.-H.;
  Hlevyack,~J.~A.; Ryu,~H.; Kee,~H.-Y.; Tamura,~N.; Chou,~M.-Y.; Hussain,~Z.;
  Mo,~S.-K.; Chiang,~T.-C. Elemental Topological Dirac Semimetal:
  $\ensuremath{\alpha}$-Sn on InSb(111). \emph{Phys. Rev. Lett.} \textbf{2017},
  \emph{118}, 146402\relax
\mciteBstWouldAddEndPuncttrue
\mciteSetBstMidEndSepPunct{\mcitedefaultmidpunct}
{\mcitedefaultendpunct}{\mcitedefaultseppunct}\relax
\EndOfBibitem
\bibitem[Li \latin{et~al.}(2022)Li, Yu, Cui, Chen, Lai, Cheng, and
  He]{10.1063/5.0084762}
Li,~M.; Yu,~J.; Cui,~G.; Chen,~Y.; Lai,~Y.; Cheng,~S.; He,~K. {Circular
  photogalvanic effect of surface states in the topological insulator
  Bi$_2$(Te$_{0.23}$Se$_{0.77}$)$_3$ nanowires grown by chemical vapor
  deposition}. \emph{J. Appl. Phys.} \textbf{2022}, \emph{131}, 113902\relax
\mciteBstWouldAddEndPuncttrue
\mciteSetBstMidEndSepPunct{\mcitedefaultmidpunct}
{\mcitedefaultendpunct}{\mcitedefaultseppunct}\relax
\EndOfBibitem
\bibitem[Zheng \latin{et~al.}(2019)Zheng, Zhang, Tong, and Du]{Zheng_2020}
Zheng,~X.; Zhang,~J.-F.; Tong,~B.; Du,~R.-R. Epitaxial growth and electronic
  properties of few-layer stanene on InSb (111). \emph{2D Materials}
  \textbf{2019}, \emph{7}, 011001\relax
\mciteBstWouldAddEndPuncttrue
\mciteSetBstMidEndSepPunct{\mcitedefaultmidpunct}
{\mcitedefaultendpunct}{\mcitedefaultseppunct}\relax
\EndOfBibitem
\bibitem[Ding \latin{et~al.}(2021)Ding, Liu, Zhang, Kalappattil, Yu, Erugu,
  Tang, Ding, Chen, and Wu]{https://doi.org/10.1002/adfm.202008411}
Ding,~J.; Liu,~C.; Zhang,~Y.; Kalappattil,~V.; Yu,~R.; Erugu,~U.; Tang,~J.;
  Ding,~H.; Chen,~H.; Wu,~M. Large damping enhancement in
  dirac-semimetal–ferromagnetic-metal layered structures caused by
  topological surface states. \emph{Adv. Funct. Mater.} \textbf{2021},
  \emph{31}, 2008411\relax
\mciteBstWouldAddEndPuncttrue
\mciteSetBstMidEndSepPunct{\mcitedefaultmidpunct}
{\mcitedefaultendpunct}{\mcitedefaultseppunct}\relax
\EndOfBibitem
\bibitem[Ding \latin{et~al.}(2021)Ding, Liu, Kalappattil, Zhang, Mosendz,
  Erugu, Yu, Tian, DeMann, Field, Yang, Ding, Tang, Terris, Fert, Chen, and
  Wu]{https://doi.org/10.1002/adma.202005909}
Ding,~J. \latin{et~al.}  Switching of a Magnet by Spin-Orbit Torque from a
  Topological Dirac Semimetal. \emph{Adv. Mater.} \textbf{2021}, \emph{33},
  2005909\relax
\mciteBstWouldAddEndPuncttrue
\mciteSetBstMidEndSepPunct{\mcitedefaultmidpunct}
{\mcitedefaultendpunct}{\mcitedefaultseppunct}\relax
\EndOfBibitem
\bibitem[Alam \latin{et~al.}(2024)Alam, Kazakov, Ahmad, Islam, Xue, and
  Matusiak]{PhysRevB.109.245135}
Alam,~M.~S.; Kazakov,~A.; Ahmad,~M.; Islam,~R.; Xue,~F.; Matusiak,~M. Quantum
  transport properties of the topological Dirac semimetal
  \ensuremath{\alpha}-Sn. \emph{Phys. Rev. B} \textbf{2024}, \emph{109},
  245135\relax
\mciteBstWouldAddEndPuncttrue
\mciteSetBstMidEndSepPunct{\mcitedefaultmidpunct}
{\mcitedefaultendpunct}{\mcitedefaultseppunct}\relax
\EndOfBibitem
\bibitem[Falson \latin{et~al.}(2020)Falson, Xu, Liao, Zang, Zhu, Wang, Zhang,
  Liu, Duan, He, Liu, Smet, Zhang, and Xue]{10.1126/science.aax3873}
Falson,~J.; Xu,~Y.; Liao,~M.; Zang,~Y.; Zhu,~K.; Wang,~C.; Zhang,~Z.; Liu,~H.;
  Duan,~W.; He,~K.; Liu,~H.; Smet,~J.~H.; Zhang,~D.; Xue,~Q.-K. Type-II Ising
  pairing in few-layer stanene. \emph{Science} \textbf{2020}, \emph{367},
  1454--1457\relax
\mciteBstWouldAddEndPuncttrue
\mciteSetBstMidEndSepPunct{\mcitedefaultmidpunct}
{\mcitedefaultendpunct}{\mcitedefaultseppunct}\relax
\EndOfBibitem
\bibitem[Engel \latin{et~al.}(2024)Engel, Dempsey, Inbar, Dong, Nishihaya,
  Chang, Fedorov, Hashimoto, Lu, and Palmstr\o{}m]{PhysRevMaterials.8.044202}
Engel,~A.~N.; Dempsey,~C.~P.; Inbar,~H.~S.; Dong,~J.~T.; Nishihaya,~S.;
  Chang,~Y.; Fedorov,~A.~V.; Hashimoto,~M.; Lu,~D.; Palmstr\o{}m,~C.~J. Growth
  and characterization of $\ensuremath{\alpha}$-Sn thin films on In- and
  Sb-rich reconstructions of InSb(001). \emph{Phys. Rev. Mater.} \textbf{2024},
  \emph{8}, 044202\relax
\mciteBstWouldAddEndPuncttrue
\mciteSetBstMidEndSepPunct{\mcitedefaultmidpunct}
{\mcitedefaultendpunct}{\mcitedefaultseppunct}\relax
\EndOfBibitem
\bibitem[Wu \latin{et~al.}(2014)Wu, Shan, and Yan]{PhysRevLett.113.256401}
Wu,~S.-C.; Shan,~G.; Yan,~B. Prediction of Near-Room-Temperature Quantum
  Anomalous Hall Effect on Honeycomb Materials. \emph{Phys. Rev. Lett.}
  \textbf{2014}, \emph{113}, 256401\relax
\mciteBstWouldAddEndPuncttrue
\mciteSetBstMidEndSepPunct{\mcitedefaultmidpunct}
{\mcitedefaultendpunct}{\mcitedefaultseppunct}\relax
\EndOfBibitem
\bibitem[Chenxiao and Jia(2020)Chenxiao, and Jia]{10.1007/s11467-020-0965-5}
Chenxiao,~Z.; Jia,~J.-F. Stanene: A good platform for topological insulator and
  topological superconductor. \emph{Front. Phys} \textbf{2020}, \emph{15}\relax
\mciteBstWouldAddEndPuncttrue
\mciteSetBstMidEndSepPunct{\mcitedefaultmidpunct}
{\mcitedefaultendpunct}{\mcitedefaultseppunct}\relax
\EndOfBibitem
\bibitem[Xu \latin{et~al.}(2015)Xu, Tang, and Zhang]{PhysRevB.92.081112}
Xu,~Y.; Tang,~P.; Zhang,~S.-C. Large-gap quantum spin Hall states in decorated
  stanene grown on a substrate. \emph{Phys. Rev. B} \textbf{2015}, \emph{92},
  081112\relax
\mciteBstWouldAddEndPuncttrue
\mciteSetBstMidEndSepPunct{\mcitedefaultmidpunct}
{\mcitedefaultendpunct}{\mcitedefaultseppunct}\relax
\EndOfBibitem
\bibitem[Barfuss \latin{et~al.}(2013)Barfuss, Dudy, Scholz, Roth, H\"opfner,
  Blumenstein, Landolt, Dil, Plumb, Radovic, Bostwick, Rotenberg, Fleszar,
  Bihlmayer, Wortmann, Li, Hanke, Claessen, and
  Sch\"afer]{PhysRevLett.111.157205}
Barfuss,~A. \latin{et~al.}  Elemental Topological Insulator with Tunable Fermi
  Level: Strained $\ensuremath{\alpha}$-Sn on InSb(001). \emph{Phys. Rev.
  Lett.} \textbf{2013}, \emph{111}, 157205\relax
\mciteBstWouldAddEndPuncttrue
\mciteSetBstMidEndSepPunct{\mcitedefaultmidpunct}
{\mcitedefaultendpunct}{\mcitedefaultseppunct}\relax
\EndOfBibitem
\bibitem[Polaczyński \latin{et~al.}(2024)Polaczyński, Krizman, Kazakov,
  Turowski, Ortiz, Rudniewski, Wojciechowski, Dłużewski, Aleszkiewicz,
  Zaleszczyk, Kurowska, Muhammad, Rosmus, Olszowska, {de Vaulchier}, Guldner,
  Wojtowicz, and Volobuev]{POLACZYNSKI2024135}
Polaczyński,~J. \latin{et~al.}  3D topological semimetal phases of strained
  \ensuremath{\alpha}-Sn on insulating substrate. \emph{Mater. Today}
  \textbf{2024}, \emph{75}, 135--148\relax
\mciteBstWouldAddEndPuncttrue
\mciteSetBstMidEndSepPunct{\mcitedefaultmidpunct}
{\mcitedefaultendpunct}{\mcitedefaultseppunct}\relax
\EndOfBibitem
\bibitem[Ding \latin{et~al.}(2022)Ding, Li, Zhou, Lu, and
  Chen]{10.1063/5.0098585}
Ding,~Y.; Li,~C.; Zhou,~J.; Lu,~H.; Chen,~Y.-F. Transport evidence of the
  spin-polarized topological surface states of $\ensuremath{\alpha}$-Sn grown
  on CdTe by molecular beam epitaxy. \emph{Appl. Phys. Lett.} \textbf{2022},
  \emph{121}, 093102\relax
\mciteBstWouldAddEndPuncttrue
\mciteSetBstMidEndSepPunct{\mcitedefaultmidpunct}
{\mcitedefaultendpunct}{\mcitedefaultseppunct}\relax
\EndOfBibitem
\bibitem[Huang and Liu(2017)Huang, and Liu]{Huang2017}
Huang,~H.; Liu,~F. Tensile strained gray tin: {D}irac semimetal for observing
  negative magnetoresistance with Shubnikov–de {H}aas oscillations.
  \emph{Phys. Rev. B} \textbf{2017}, \emph{95}, 201101\relax
\mciteBstWouldAddEndPuncttrue
\mciteSetBstMidEndSepPunct{\mcitedefaultmidpunct}
{\mcitedefaultendpunct}{\mcitedefaultseppunct}\relax
\EndOfBibitem
\bibitem[Zang \latin{et~al.}(2018)Zang, Jiang, Gong, Guan, Liu, Liao, Zhu, Li,
  Wang, Li, Song, Zhang, Xu, He, Ma, Zhang, and Xue]{10.1002/adfm.201802723}
Zang,~Y. \latin{et~al.}  Realizing an Epitaxial Decorated Stanene with an
  Insulating Bandgap. \emph{Adv. Funct. Mater.} \textbf{2018}, \emph{28},
  1802723\relax
\mciteBstWouldAddEndPuncttrue
\mciteSetBstMidEndSepPunct{\mcitedefaultmidpunct}
{\mcitedefaultendpunct}{\mcitedefaultseppunct}\relax
\EndOfBibitem
\bibitem[Basnet \latin{et~al.}(2024)Basnet, Upreti, McCarthy, Ju, McMinn,
  Sharma, Zhang, and Hu]{10.1116/6.0003564}
Basnet,~R.; Upreti,~D.; McCarthy,~T.~T.; Ju,~Z.; McMinn,~A.~M.; Sharma,~M.~M.;
  Zhang,~Y.-H.; Hu,~J. Magneto-transport study on Sn-rich Sn$_{1-x}$Ge$_x$ thin
  films enabled by CdTe buffer layer. \emph{J VAC SCI TECHNOL B} \textbf{2024},
  \emph{42}, 042210\relax
\mciteBstWouldAddEndPuncttrue
\mciteSetBstMidEndSepPunct{\mcitedefaultmidpunct}
{\mcitedefaultendpunct}{\mcitedefaultseppunct}\relax
\EndOfBibitem
\bibitem[Chen \latin{et~al.}(2022)Chen, Lin, Lien, Huang, Cheng, Lin, Hsu,
  Chang, Cheng, Hong, and Kwo]{PhysRevB.105.075109}
Chen,~K. H.~M.; Lin,~K.~Y.; Lien,~S.~W.; Huang,~S.~W.; Cheng,~C.~K.;
  Lin,~H.~Y.; Hsu,~C.-H.; Chang,~T.-R.; Cheng,~C.-M.; Hong,~M.; Kwo,~J.
  Thickness-dependent topological phase transition and Rashba-like preformed
  topological surface states of \ensuremath{\alpha}-Sn(001) thin films on
  InSb(001). \emph{Phys. Rev. B} \textbf{2022}, \emph{105}, 075109\relax
\mciteBstWouldAddEndPuncttrue
\mciteSetBstMidEndSepPunct{\mcitedefaultmidpunct}
{\mcitedefaultendpunct}{\mcitedefaultseppunct}\relax
\EndOfBibitem
\bibitem[Jardine \latin{et~al.}(2023)Jardine, Dardzinski, Yu, Purkayastha,
  Chen, Chang, Engel, Strocov, Hocevar, Palmstrom, Frolov, and
  Marom]{10.1021/acsami.3c00323}
Jardine,~M. J.~A.; Dardzinski,~D.; Yu,~M.; Purkayastha,~A.; Chen,~A.-H.;
  Chang,~Y.-H.; Engel,~A.; Strocov,~V.~N.; Hocevar,~M.; Palmstrom,~C.;
  Frolov,~S.~M.; Marom,~N. First-Principles Assessment of CdTe as a Tunnel
  Barrier at the $\alpha$-Sn/InSb Interface. \emph{ACS Appl. Mater. Interfaces}
  \textbf{2023}, \emph{15}, 16288--16298, PMID: 36940162\relax
\mciteBstWouldAddEndPuncttrue
\mciteSetBstMidEndSepPunct{\mcitedefaultmidpunct}
{\mcitedefaultendpunct}{\mcitedefaultseppunct}\relax
\EndOfBibitem
\bibitem[Fu and Kane(2007)Fu, and Kane]{PhysRevB.76.045302}
Fu,~L.; Kane,~C.~L. Topological insulators with inversion symmetry. \emph{Phys.
  Rev. B} \textbf{2007}, \emph{76}, 045302\relax
\mciteBstWouldAddEndPuncttrue
\mciteSetBstMidEndSepPunct{\mcitedefaultmidpunct}
{\mcitedefaultendpunct}{\mcitedefaultseppunct}\relax
\EndOfBibitem
\bibitem[Si \latin{et~al.}(2020)Si, Yao, Jiang, Li, Zhou, Ji, Huang, Li, and
  Niu]{doi:10.1021/acs.jpclett.9b03538}
Si,~N.; Yao,~Q.; Jiang,~Y.; Li,~H.; Zhou,~D.; Ji,~Q.; Huang,~H.; Li,~H.;
  Niu,~T. Recent Advances in Tin: From Two-Dimensional Quantum Spin Hall
  Insulator to Bulk Dirac Semimetal. \emph{J. Phys. Chem. Lett.} \textbf{2020},
  \emph{11}, 1317--1329\relax
\mciteBstWouldAddEndPuncttrue
\mciteSetBstMidEndSepPunct{\mcitedefaultmidpunct}
{\mcitedefaultendpunct}{\mcitedefaultseppunct}\relax
\EndOfBibitem
\bibitem[Mciver \latin{et~al.}(2011)Mciver, Hsieh, Steinberg, Jarillo-Herrero,
  and Gedik]{10.1038/nnano.2011.214}
Mciver,~J.; Hsieh,~D.; Steinberg,~H.; Jarillo-Herrero,~P.; Gedik,~N. Control
  over topological insulator photocurrents with light polarization. \emph{Nat.
  Nanotechnol.} \textbf{2011}, \emph{7}, 96--100\relax
\mciteBstWouldAddEndPuncttrue
\mciteSetBstMidEndSepPunct{\mcitedefaultmidpunct}
{\mcitedefaultendpunct}{\mcitedefaultseppunct}\relax
\EndOfBibitem
\bibitem[Ding \latin{et~al.}(2021)Ding, Yao, Yuan, Li, Lu, Lu, and
  Chen]{10.1116/6.0000756}
Ding,~Y.; Yao,~J.; Yuan,~Z.; Li,~C.; Lu,~M.-H.; Lu,~H.; Chen,~Y.-F. Multiple
  carrier transport in high-quality $\alpha$-Sn films grown on CdTe (001) by
  molecular beam epitaxy. \emph{J VAC SCI TECHNOL A} \textbf{2021}, \emph{39},
  033408\relax
\mciteBstWouldAddEndPuncttrue
\mciteSetBstMidEndSepPunct{\mcitedefaultmidpunct}
{\mcitedefaultendpunct}{\mcitedefaultseppunct}\relax
\EndOfBibitem
\bibitem[Plank \latin{et~al.}(2016)Plank, Golub, Bauer, Bel'kov, Herrmann,
  Olbrich, Eschbach, Plucinski, Schneider, Kampmeier, Lanius, Mussler,
  Gr\"utzmacher, and Ganichev]{PhysRevB.93.125434}
Plank,~H.; Golub,~L.~E.; Bauer,~S.; Bel'kov,~V.~V.; Herrmann,~T.; Olbrich,~P.;
  Eschbach,~M.; Plucinski,~L.; Schneider,~C.~M.; Kampmeier,~J.; Lanius,~M.;
  Mussler,~G.; Gr\"utzmacher,~D.; Ganichev,~S.~D. Photon drag effect in
  $({\mathrm{Bi}}_{1\ensuremath{-}x}{\mathrm{Sb}}_{x}){}_{2}{\mathrm{Te}}_{3}$
  three-dimensional topological insulators. \emph{Phys. Rev. B} \textbf{2016},
  \emph{93}, 125434\relax
\mciteBstWouldAddEndPuncttrue
\mciteSetBstMidEndSepPunct{\mcitedefaultmidpunct}
{\mcitedefaultendpunct}{\mcitedefaultseppunct}\relax
\EndOfBibitem
\bibitem[Hamh \latin{et~al.}(2016)Hamh, Park, Jerng, Jeon, Chun, and
  Lee]{PhysRevB.94.161405}
Hamh,~S.~Y.; Park,~S.-H.; Jerng,~S.-K.; Jeon,~J.~H.; Chun,~S.-H.; Lee,~J.~S.
  Helicity-dependent photocurrent in a ${\mathrm{Bi}}_{2}{\mathrm{Se}}_{3}$
  thin film probed by terahertz emission spectroscopy. \emph{Phys. Rev. B}
  \textbf{2016}, \emph{94}, 161405\relax
\mciteBstWouldAddEndPuncttrue
\mciteSetBstMidEndSepPunct{\mcitedefaultmidpunct}
{\mcitedefaultendpunct}{\mcitedefaultseppunct}\relax
\EndOfBibitem
\bibitem[Yu \latin{et~al.}(2019)Yu, Zhu, Zeng, Chen, Chen, Liu, Yin, Cheng,
  Lai, Huang, He, and Xue]{PhysRevB.100.235108}
Yu,~J.; Zhu,~K.; Zeng,~X.; Chen,~L.; Chen,~Y.; Liu,~Y.; Yin,~C.; Cheng,~S.;
  Lai,~Y.; Huang,~J.; He,~K.; Xue,~Q. Helicity-dependent photocurrent of the
  top and bottom Dirac surface states of epitaxial thin films of
  three-dimensional topological insulators
  ${\mathrm{Sb}}_{2}{\mathrm{Te}}_{3}$. \emph{Phys. Rev. B} \textbf{2019},
  \emph{100}, 235108\relax
\mciteBstWouldAddEndPuncttrue
\mciteSetBstMidEndSepPunct{\mcitedefaultmidpunct}
{\mcitedefaultendpunct}{\mcitedefaultseppunct}\relax
\EndOfBibitem
\bibitem[Jochym \latin{et~al.}(2022)Jochym, Kremer, \L{}a\ifmmode~\dot{z}\else
  \.{z}\fi{}ewski, Ptok, Piekarz, Br\"ucher, and Ole\ifmmode~\acute{s}\else
  \'{s}\fi{}]{PhysRevMaterials.6.113601}
Jochym,~P.~T.; Kremer,~R.~K.; \L{}a\ifmmode~\dot{z}\else \.{z}\fi{}ewski,~J.;
  Ptok,~A.; Piekarz,~P.; Br\"ucher,~E.; Ole\ifmmode~\acute{s}\else
  \'{s}\fi{},~A.~M. Influence of anharmonicity on the negative thermal
  expansion of $\ensuremath{\alpha}\text{\ensuremath{-}}\mathrm{Sn}$.
  \emph{Phys. Rev. Mater.} \textbf{2022}, \emph{6}, 113601\relax
\mciteBstWouldAddEndPuncttrue
\mciteSetBstMidEndSepPunct{\mcitedefaultmidpunct}
{\mcitedefaultendpunct}{\mcitedefaultseppunct}\relax
\EndOfBibitem
\bibitem[van Veen \latin{et~al.}(2025)van Veen, K\"olling, de~Wit, Metsch,
  Rosenbach, Li, and Brinkman]{d5xc-bvmx}
van Veen,~F.; K\"olling,~S.; de~Wit,~S.~R.; Metsch,~R.; Rosenbach,~D.; Li,~C.;
  Brinkman,~A. Observation of the surface hybridization gap in the electrical
  transport properties of the ultrathin topological insulator
  ${({\mathrm{Bi}}_{1\ensuremath{-}x}{\mathrm{Sb}}_{x})}_{2}{\mathrm{Te}}_{3}$.
  \emph{Phys. Rev. B} \textbf{2025}, \emph{112}, 045425\relax
\mciteBstWouldAddEndPuncttrue
\mciteSetBstMidEndSepPunct{\mcitedefaultmidpunct}
{\mcitedefaultendpunct}{\mcitedefaultseppunct}\relax
\EndOfBibitem
\bibitem[Zhang \latin{et~al.}(2010)Zhang, He, Chang, Song, Wang, Chen, Jia,
  Fang, Dai, Shan, Shen, Niu, Qi, Zhang, Ma, and Xue]{WOS:000280559300013}
Zhang,~Y. \latin{et~al.}  Crossover of the three-dimensional topological
  insulator Bi2Se3 to the two-dimensional limit. \emph{Nat. Phys.}
  \textbf{2010}, \emph{6}, 584--588\relax
\mciteBstWouldAddEndPuncttrue
\mciteSetBstMidEndSepPunct{\mcitedefaultmidpunct}
{\mcitedefaultendpunct}{\mcitedefaultseppunct}\relax
\EndOfBibitem
\end{mcitethebibliography}

\end{document}